\documentclass{article}
\usepackage{arxiv}
\usepackage[utf8]{inputenc} 
\usepackage[T1]{fontenc}    
\usepackage{hyperref}       
\usepackage{url}            
\usepackage{booktabs}       
\usepackage{amsfonts}       
\usepackage{nicefrac}       
\usepackage{microtype}      
\usepackage{lipsum}		
\usepackage{graphicx}
\usepackage{natbib}
\usepackage{doi}
\usepackage{amsmath,amssymb,amsfonts}
\usepackage{multicol,multirow}

\title{Modeling Snow  on Sea Ice using Physics Guided Machine Learning}

\author{ 
    \href{https://orcid.org/0000-0001-9411-7450}{\includegraphics[scale=0.06]{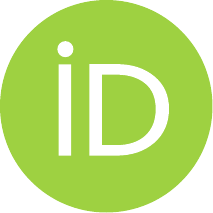}\hspace{1mm}Ayush Prasad}\thanks{Corresponding author. Email: ayush.prasad@fmi.fi} \\
    Finnish Meteorological Institute \\
    Helsinki, Finland
    \And
    \href{https://orcid.org/0000-0002-1878-3731}{\includegraphics[scale=0.06]{orcid.pdf}\hspace{1mm}Ioanna Merkouriadi} \\
    Finnish Meteorological Institute \\
    Helsinki, Finland
    \And
    \href{https://orcid.org/0000-0002-7591-3504}{\includegraphics[scale=0.06]{orcid.pdf}\hspace{1mm}Aleksi Nummelin} \\
    Finnish Meteorological Institute \\
    Helsinki, Finland
}

\renewcommand{\shorttitle}{Modeling Snow  on Sea Ice using Physics Guided Machine Learning}

\hypersetup{
pdftitle={Modeling Snow  on Sea Ice using Physics Guided Machine Learning},
}

\begin{document}
\maketitle

\begin{abstract}
Snow is a crucial element of the sea ice system, affecting the sea ice growth and decay due to its low thermal conductivity and high albedo. Despite its importance, present-day climate models have a very idealized representation of snow, often including just 1-layer thermodynamics, omitting several processes that shape its properties. Even though sophisticated snow process models exist, they tend to be excluded in climate modeling due to their prohibitive computational costs. For example, SnowModel is a numerical snow process model developed to simulate the evolution of snow depth and density, blowing-snow redistribution and sublimation, snow grain size, and thermal conductivity, in a spatially distributed, multi-layer snowpack framework. SnowModel can simulate snow distributions on sea ice floes in high spatial (1-m horizontal grid) and temporal (1-hour time step) resolution. However, for simulations spanning over large regions, such as the Arctic Ocean, high-resolution runs face challenges of slow processing speeds and the need for large computational resources. To address these common issues in high-resolution numerical modeling, data-driven emulators are often used. However, these emulators have their caveats, primarily a lack of generalizability and inconsistency with physical laws. In our study, we address these challenges by using a physics-guided approach in developing our emulator. By integrating physical laws that govern changes in snow density due to compaction, we aim to create an emulator that is efficient while also adhering to essential physical principles. We evaluated this approach by comparing three machine learning models: Long Short-Term Memory (LSTM), Physics-Guided LSTM and Random Forest, across five distinct Arctic regions. Our evaluations indicate that all models achieved high accuracy, with the Physics-Guided LSTM model demonstrating the most promising results in terms of accuracy and generalizability. Our approach offers a computationally faster way to emulate the SnowModel with high fidelity and a speedup of over 9000 times.
\end{abstract}


\section{Introduction}
Snow is a critical component of the Arctic sea ice system, with unique thermal and optical properties that control sea ice thermodynamics. Snow has low thermal conductivity values that hinder sea ice growth in winter \cite{Sturm2002,Macfarlane2023}, and high albedo that delays the sea ice from melting in summer. Both the thermal and optical properties of snow are affected by the state of the snowpack, i.e., the snow depth, density, and the snow grain size and type of different snow layers. Information on the spatial distribution and temporal evolution of the snowpack states is important for both process modeling and remote sensing applications. A big challenge associated with scaling is that snowpack states demonstrate large spatial variations at scales of a few meters that are beyond the coarse spatial resolution of both satellite observations and large-scale sea ice models.

SnowModel, is a numerical process model that simulates the evolution and spatial distribution of snow physical properties and structural characteristics in a multi-layer snowpack framework \cite{Liston2006}. It is applicable to any environment experiencing snow, including sea ice applications \cite{Liston2018}. SnowModel is a collection of modeling tools designed to be versatile in terms of spatial domain (from regional to global applications), temporal domain (for past, current, and future climates), spatial resolution (e.g., 1-m to 100-km), and temporal resolution (e.g. 1-hour to daily). Meteorological information is required as an input to drive the SnowModel. It specifically requires meteorological station, or atmospheric datasets, that include air temperature, relative humidity, precipitation, wind speed, and wind direction. Topography information is also required for simulating snow distributions, because in SnowModel snow is blown and redistributed by the wind, accumulating in the lee side of wind obstruction features.

A main limitation of physical numerical models, like SnowModel, lies in their high computational requirements which makes them less feasible for some practical applications due to the high resources required. SnowModel specifically, is written in Fortran 77 and has no model components that can be run in parallel making its performance even slower. On the other hand, machine learning (ML) has shown potential in enhancing climate models, particularly in replacing or supplementing components of traditional models. This is evident in applications such as simulating cloud convection using random forests and deep learning models \cite{beucler2023machine}, and in emulating ocean general circulation models through neural networks \cite{agarwal2021comparison}. In the cryosphere, recent developments include the creation of emulators for Ice Sheet Models to project sea level rise using Gaussian processes \cite{berdahl2021statistical} and Long Short-Term Memory (LSTM) networks \cite{van2023variational}. Similarly, emulators have been used for improving the paramterizations in sea-ice models \cite{driscoll2024parameter, gregory2024machine}.  

Despite the progress made with machine learning, data-driven emulators face their own set of challenges. First, there is the concern of generalizability: these models often struggle to extend their applicability beyond the conditions and data they were trained on \cite{jebeile2023machine}. Second, the opacity of ML models poses a significant issue. Often referred to as "black box" models, they lack transparency, offering little to no insight into the reasoning behind their predictions\cite{mcgovern2022we}. Third, a critical limitation is their inability to adhere to fundamental physical laws, including the conservation of mass, which is an important principle in process-guided models. \cite{kashinath2021physics, sudakow2022statistical}. To tackle these challenges, physics-guided machine learning has recently emerged as a promising approach. This involves incorporating physical laws within the ML framework. It can be achieved by softly constraining the loss function of the model, thereby guiding it to make predictions that are not only accurate but also consistent with physical principles \cite{kashinath2021physics}. In broader climate science, recent studies have demonstrated that physics-guided machine learning models exhibit robustness and have a higher adherence to physical principles. Additionally, these models demonstrate increased data efficiency, requiring fewer data points for training compared to traditional machine learning models that do not incorporate physical constraints. \cite{kashinath2021physics, beucler2021enforcing}. In our work, we develop and evaluate machine learning-based emulators of SnowModel for modeling snow on sea ice. Our key contributions are: 
\begin{itemize}
    \item We develop and compare machine learning architectures for predicting the density of snow on sea ice. 
    \item We explore the use of physics-guided loss functions to constrain the emulator using the underlying physics of snowpack formation.
\end{itemize}

\section[Methods]{Methods}

\subsection{Study Region and Dataset construction}
\begin{figure}[h]
\centering
\includegraphics[width=3.3in, height=3in]{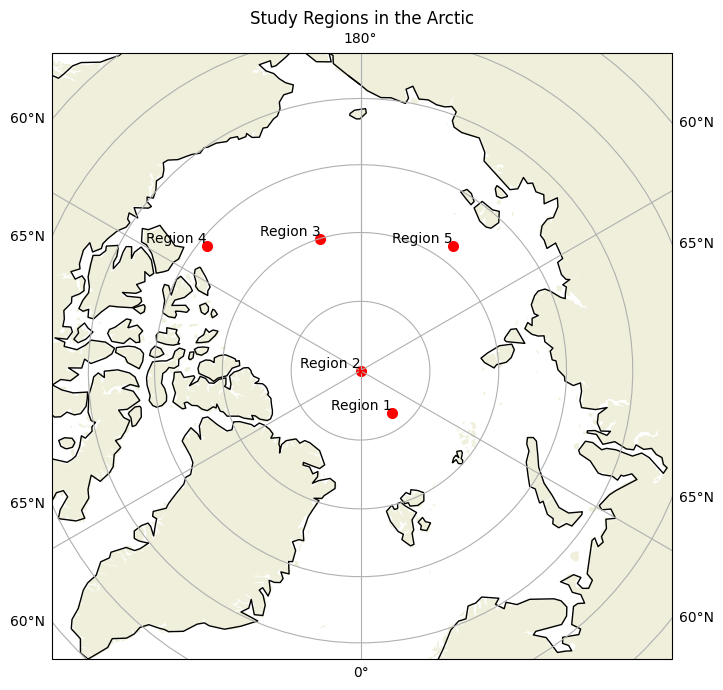}
\caption{Arctic Study Regions}
\label{fig:1}
\end{figure}

We utilized the SnowModel to generate a comprehensive dataset covering the snow dynamics in various regions in the Arctic. Due to the lack of availability of high resolution sea-ice topography data,  we synthetically created ice topographic features in these regions by generating random sea-ice ridges based on the minimum, maxium and standard deviation of the sea ice ridges described in \cite{sudom2011}. The dataset for the emulator was created by running the SnowModel across five distinct Arctic regions (Figure \ref{fig:1}) over ten years. The selected regions spanned over different Arctic Seas that are influenced by a wide range of atmospheric conditions (Central Arctic, Beaufort, Chuckhi, Laptev and Barents Seas), resulting in a broad variety of snowpack characteristics. 

\subsection{SnowModel Configuration}
We used NASA's Modern Era Retrospective Analysis for Research and Application Version 2 \cite[MERRA-2; ][]{merra2} as meteorological forcing to SnowModel. Specifically, SnowModel was forced with 10-m wind speed and direction, 2-m air temperature and relative humidity, and total water-equivalent precipitation from MERRA-2 (Table \ref{tab:variables_snow_density}). The simulations began in August 2010 and ran through August 2020. Temporal resolution was 3 hours to capture diurnal variations of the snow properties. The outputs were saved daily, at the end of each day.

\begin{table}[h]
\centering
\caption{Variables driving SnowModel and inputs for emulators}
\label{tab:variables_snow_density}
\begin{tabular}{ll}
\hline
\textbf{Source} & \textbf{Feature} \\
\hline
\multirow{4}{*}{MERRA2} & Precipitation \\
& 2m Air Temperature \\
& Wind Speed \\
& Relative humidity \\
& Day of Year \\
Synthetically Generated & Topography\\
\hline
\end{tabular}
\end{table}

\subsection{LSTM Architecture}
Long Short-Term Memory (LSTM) networks are a special class of Recurrent Neural Networks (RNNs) that are designed to address the challenge of learning long-term dependencies in sequential data \cite{hochreiter1997long}. Traditional RNNs often struggle with the vanishing or exploding gradient problem, which hampers their ability to retain information over extended sequences \cite{bengio1994learning}. LSTMs mitigate this issue through a unique architecture consisting of memory cells and multiple gates. Each LSTM unit contains three gates: input, forget, and output. The input gate controls the influx of new information into the cell state, the forget gate manages the retention or removal of information, and the output gate influences the contribution of the cell state to the output. This gating mechanism is crucial for preserving information across long sequences, allowing LSTMs to effectively tackle tasks involving sequential data, such as time-series data. In our model, the input layer accommodates the five primary features. This is followed by two LSTM layers with 128 units each, for capturing the sequential patterns in the data. Dropout layers, set at a rate of 0.25, are used after each LSTM layer to prevent overfitting. A dense layer with 64 units featuring a ReLU activation function is integrated, succeeded by a dropout layer at a rate of 0.25. The final output layer is structured to predict snow density as well as two additional variables snow depth and snow temperature. We use these additional variables while constructing the physics-constrained loss function for the LSTM. The models were implemented in Python using PyTorch \cite{paszke2019pytorch}.

\subsection{Physics Guided LSTM}
We developed a loss function to incorporate the physics of snowpack formation in the LSTM model. As snow accumulates, the lower layers of the snowpack are subjected to increasing pressure due to the weight of the snow above. This pressure causes the snow grains in the lower layers to compact or press closer together. As a result, the density of the snow in these lower layers increases. This process of density increase due to compaction is included in SnowModel's SnowPack component and it provides key equations explaining the snow compaction dynamics \cite{liston2006distributed}. The equation for density increase due to compaction is given by:

\begin{equation}
\frac{\partial\rho_s}{\partial t} = A_1 \cdot h_w^{*} \cdot \frac{\rho_s \exp(-B(T_f - T_s))}{\exp(-A_2 \rho_s)}
\label{eq:density_compaction}
\end{equation}

where \( \rho_s \) is the snow density (kg/m\(^3\)), \( t \) represents time (s), \(  h_w^{*}  = \frac{1}{2}  h_w\)  is the weight of snow
defined as half of the snow water equivalent $h_w$, \( T_f \) is the freezing temperature of water (K), and \( T_s \) is the average snow-ground interface and surface temperatures (K). Constants \( A_1 \), \( A_2 \), and \( B \) are derived from empirical studies: \( A_1 = 0.0013 \) m\(^{-1}\) s\(^{-1}\), \( A_2 = 0.021 \) m\(^3\) kg\(^{-1}\), and \( B = 0.08 \) K\(^{-1}\).

The snow water equivalent, \( h_w \) (m), is defined as:

\begin{equation}
h_w = \frac{\rho_{s,\text{phys}}}{\rho_{w}}\cdot\zeta_{s,\text{pred}}
\label{eq:snow_water_equivalent}
\end{equation} 

here \( \rho_w \) is the density of water (kg/m\(^3\)), and \( \zeta_s \) is the snow depth (m). \\

\noindent
In other words,

\begin{equation}
\rho_{s,\text{phys}}(t+\Delta t) = \rho_{s,\text{phys}}(t) + 0.5 \cdot \Delta t \cdot A_1 \cdot \frac{\rho_{s,\text{pred}}(t)}{\rho_w} \cdot \zeta_{s,\text{pred}}(t) \cdot \frac{\rho_{s,\text{pred}}(t) \exp(-B(T_f - T_{s,\text{pred}}(t)))}{\exp(-A_2 \rho_{s,\text{pred}}(t))}
\end{equation}

To align the LSTM model's predictions with these physical principles, we constructed a physics-informed loss function, \( L_{\text{physics}} \). This loss function consists of the mean squared errors (MSE) between the model's predicted values for snow density, depth, and temperature, and their corresponding values from the SnowPack equations.

The combined loss function is therefore expressed as:

\begin{equation}
L_{\text{total}} = L_{\text{LSTM}} + \lambda L_{\text{physics}}
\label{eq:combined_loss_function}
\end{equation}

where \( L_{\text{LSTM}} \) represents the initial loss from the LSTM model predictions, and \( \lambda \) is a regularization parameter that balances the contribution of the physics-informed loss in the overall objective function. This objective function was then used in the LSTM architecture described in section 2.3.

\begin{equation}
L_{\text{total}} = L_{\text{LSTM}} + \lambda \left[ \text{MSE}(\rho_{s,\text{pred}}, \rho_{s,\text{phys}}) \right]
\end{equation}

\subsubsection{Training Details}
In both the LSTM architectures, we used Adam optimizer, with a learning rate of 0.001, for training the models for 150 epochs with a batch size of 32. We used an early stopping mechanism, considering the validation loss after 10 epochs, to prevent overfitting. We selected model hyperparameters (number of layers, dropout, batch size and epochs) by random search with 5-fold cross-validation. We gave the random search a budget of 50 iterations for each model.\\

\subsection{Random Forest}
We chose Random Forest to compare our LSTM-based emulators. Random Forest \cite{breiman2001random} is an ensemble learning method used for classification and regression. It operates by constructing multiple decision trees at training time and outputting the class that is the mode of the classes (classification) or mean prediction (regression) of the individual trees. This method is particularly effective for prediction tasks because it reduces the risk of overfitting by averaging multiple trees, thus improving the model's generalizability. We used Random Forest to predict snow density. The Random Forest model was configured with 500 trees.

\section{Results}
\subsection{Predictive Performance}
For training and testing our emulator models, we used a cross-validation approach, specifically opting for a leave-one-out cross-validation (LOOCV) method. In this method, data from four out of five Arctic regions were used for training, while the remaining fifth region's data served for testing purposes. This process was systematically rotated, ensuring each region was used as a test set in turn, thus providing a comprehensive assessment of the models across varied Arctic conditions. We chose the LOOCV approach as opposed to a time-based split because the distinct climatic and geographical characteristics of each Arctic region can significantly influence the model's performance. A time-based split would not adequately capture these spatial variabilities and could lead to models that are well-tuned for temporal patterns but less robust to spatial differences. LOOCV ensures that each model is tested against the unique conditions of an unseen region, enhancing our ability to generalize our findings across the different regions, despite the diverse environmental conditions present within each region. 
\\

While evaluating the models, we also calculated the daily climatology of snow density in every region. This climatology served as a baseline for us to compare and contrast the emulator models' predictions with a standard reference point that reflects typical conditions. \\

For assessing the predictive performance of our emulators  (Table \ref{tab:ModelComparison}), we consider the Root Mean Squared Error (RMSE). RMSE provides a measure of the average magnitude of the models' prediction errors, and it is defined as:

\begin{equation}
\text{RMSE} = \sqrt{\frac{1}{n}\sum_{i=1}^{n}(y_i - \hat{y}_i)^2}
\label{eq:rmse}
\end{equation}
where \( y_i \) is the observed value, \( \hat{y}_i \) is the predicted value, and \( n \) is the number of observations.  \\

\begin{figure}[h]
    \centering
    \includegraphics[width=5in]{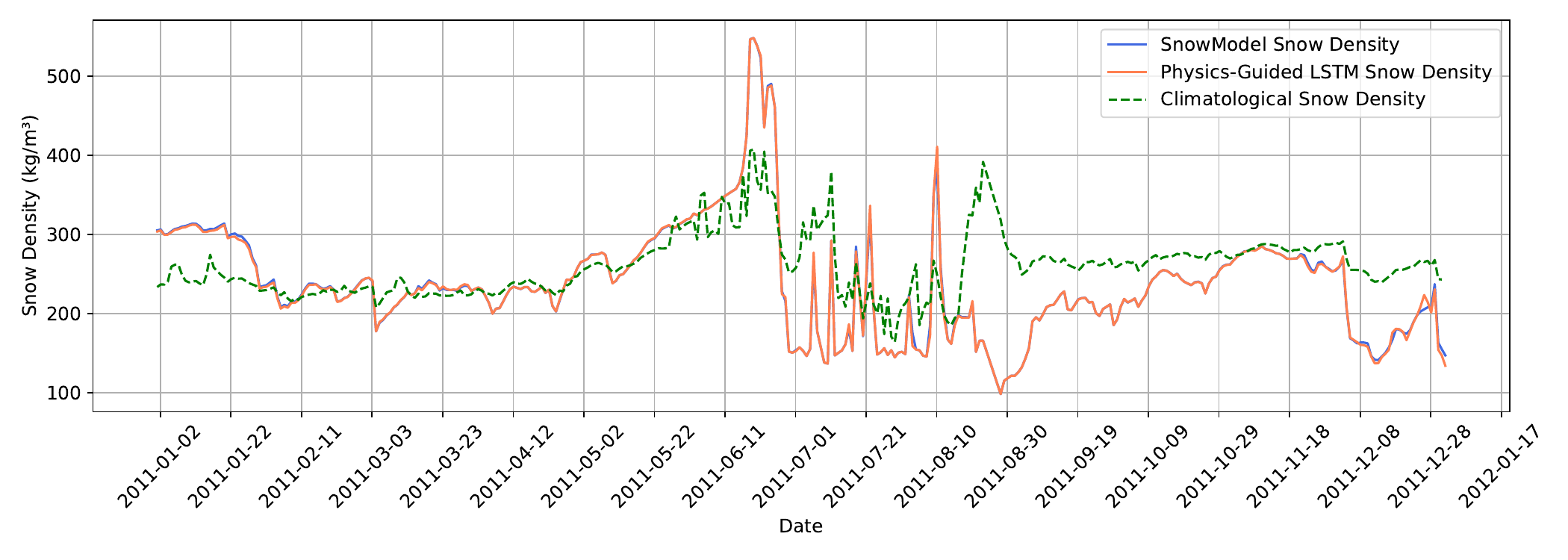}
    \caption{Predicted average snow density across all the points in region 3 by emulator compared with the snow density from SnowModel}
    \label{fig:2}
\end{figure}

\begin{figure}[h]
\centering
\includegraphics[width=5.5in]{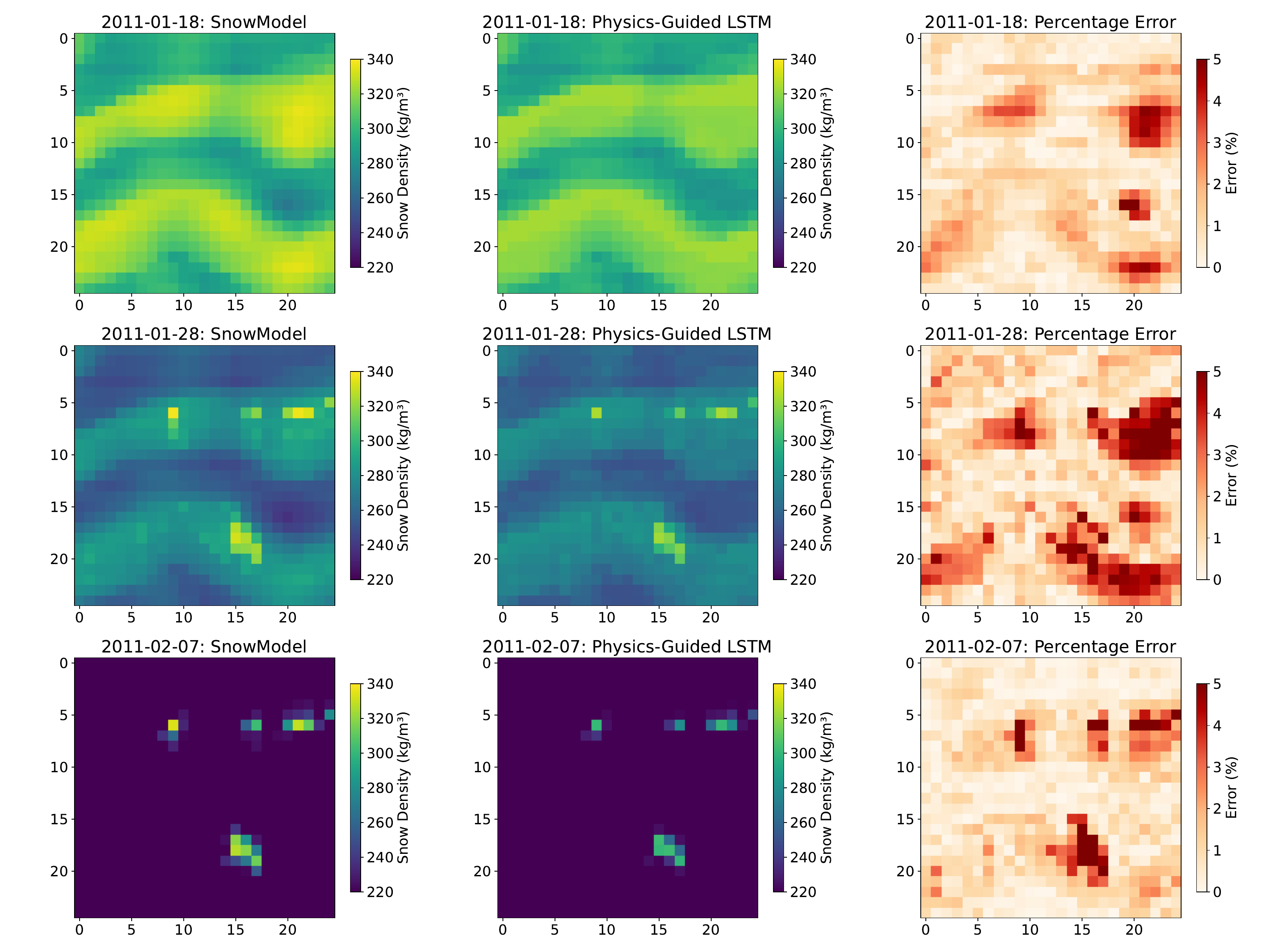}
\caption{Spatial maps showing snow density as predicted by the emulator and compared with those from the SnowModel for Region 2. The third column shows the error percentage (lower values are better), highlighting the differences between the SnowModel and the emulator.}
\label{fig:3}
\end{figure}

The results showed that all of the model's predictions closely matched the SnowModel outputs in all regions. All of the emulators performed better than the climatological baseline.  Figures \ref{fig:2} and \ref{fig:3} show a comparison of the \textit{Snow Density} as predicted by the SnowModel and Physics-Guided LSTM for Region 2. During this evaluation the emulator was trained with data from Regions 1, 3, 4 and 5 and Region 2 was the test site. The emulator's prediction aligns closely with the SnowModel which shows that the emulator can generalize to regions where it has not been trained before. The LSTM with the physics guided Loss function performed better than the LSTM with only MSE loss in four out of five regions. The overall better performance of the PG-LSTM across the other Arctic regions suggests that incorporating physical constraints through the physics-guided loss function was generally beneficial. However, every region has its own unique characteristics in terms of meteorology, topography and snowpack evolution dynamics. In Region 4, there may have been some conditions or processes that deviated from the assumptions made by the physics constraints used to guide the LSTM model. On the other hand, the Random Forest, being an ensemble of decision trees, has a more flexible model structure allowing to better perform in that region without being bound by explicit physical constraints.

\begin{table}[h]
\caption{Comparison of RMSE across the study regions for different emulator models. Lower values are better.}
\label{tab:ModelComparison}
\begin{tabular*}{\textwidth}{@{\extracolsep{\fill}}lcccc@{}}
\toprule
 & Climatology & Random Forest & LSTM & Physics-Guided LSTM \\
\midrule
Region 1  & 79.7 & 45.3 & 41.1 & \textbf{40.1} \\
Region 2  & 81.1 & 47.8 & 42.5 & \textbf{41.5} \\
Region 3  & 80.3 & 46.5 & 41.7 & \textbf{40.7} \\
Region 4  & 82.5 & \textbf{42.9} & 47.3 & 48.4 \\
Region 5  & 83.2 & 49.0 & 44.6 & \textbf{43.6} \\
\toprule
\end{tabular*}
\end{table}
\subsection{Runtime}
We conducted a runtime analysis (Table \ref{tab:runtime}) by comparing the Python runtime for all of the emulators with the Fortran runtime of the original SnowModel. The runtime measurements were conducted on a high-performance computing system with NVIDIA Volta V100 GPU and Xeon Gold 6230 CPUs with 2 x 20 cores operating at 2.1 GHz. The LSTM-based models took 8 hours to train. We measured the time required to process the data for one region of resolution 1500 x 1500 in the SnowModel.  As shown in Table \ref{tab:runtime}, the speed-up achieved by the machine learning models is significant. The Random Forest, LSTM, and Physics-Guided LSTM models demonstrated speed-ups of 2400, 13,714, and 9,931 times, respectively, compared to the SnowModel.

\begin{table}[h]
\caption{Comparison of Runtime between SnowModel and Emulators. Lower values are better.}
\label{tab:runtime}
\begin{tabular*}{\textwidth}{@{\extracolsep{\fill}}lccccc@{}}
\toprule
Model & SnowModel & Random Forest & LSTM & Physics-Guided LSTM \\
\midrule
Time (s) & 28800 & 12 & \textbf{2.1} & 2.9 \\
Speed-up & — & 2400 & \textbf{13714} & 9931 \\
\toprule
\end{tabular*}
\end{table}

\section{Feature Importances}
\begin{figure}[h]
    \centering
    \includegraphics[width=3in]{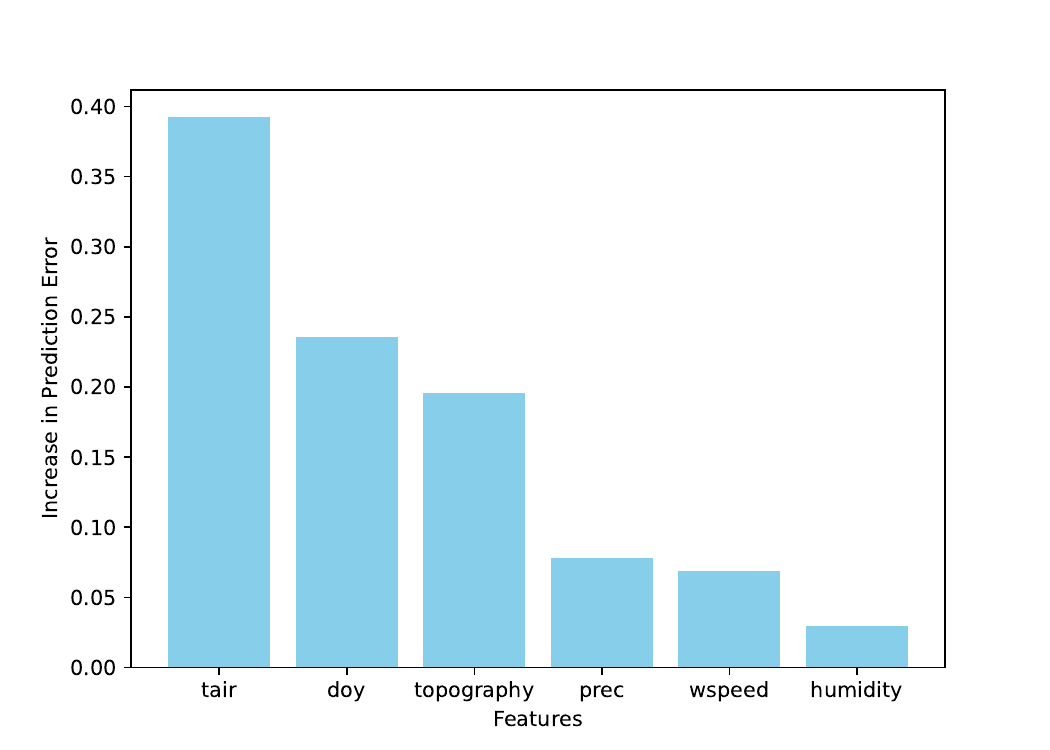}
    \caption{Feature importance derived from permutation shuffling in the Physics-guided LSTM}
    \label{fig:4}
\end{figure}

To identify the most influential variables in estimating snow density within the physics-guided LSTM model, we used permutation importance. This method involved systematically shuffling the values of each feature within the dataset for each specific Arctic site independently. We repeated this shuffling process multiple for 10 iterations to ensure reliability, and measured the effect on the model's accuracy by observing the increase in the Root Mean Squared Error (RMSE). This increase in RMSE, plotted on the y-axis in Figure \ref{fig:4}, quantitatively represents the impact of each feature's disruption on model performance. We found that air temperature and the day of the year were critical for capturing the seasonal variations in snow density, reflecting their direct influence on thermodynamics. Similarly, topography significantly affected spatial variations, influencing patterns of snow accumulation and melting due to its role in redistributing snow across the landscape. These features align with the dynamic processes in SnowModel, where snow density evolves under the influence of overlying snow weight (compaction), wind speed, sublimation of non-blowing snow, and melting processes.

\newpage

\section{Conclusion}
In this work, we explored the use of machine learning to emulate SnowModel. We developed and assessed various machine learning frameworks—Random Forest, LSTM, and Physics-Guided LSTM—to effectively emulate the SnowModel. The Physics-Guided LSTM model, in particular, showed significant improvement in prediction accuracy, as indicated by a lower RMSE. Moreover, it demonstrated a notable increase in computational speed, operating approximately 9,000 times faster than SnowModel. It is worth noting that while the LSTM model benefited from incorporating the physics of snowpack formation, due to the nature of these ‘soft’ regularizations, the constraints are not required to be strictly satisfied during runtime. In our study, we focused on emulating snow density; a similar approach could be used for emulating other variables like snow depth in SnowModel.\\

A potential direction for future research could be the integration of the emulator within a climate model to improve simulations of snow on sea ice, which are currently limited by the computational demands of numerical models like SnowModel. Additionally, the emulator itself could be developed further by exploring recent machine learning architectures like Fourier Neural Operators, which are capable of operating at arbitrary resolutions without requiring resolution-specific training. Enhancing the interpretability of these models could also be a valuable direction to ensure a clearer understanding of the model predictions and their alignment with physical processes.

\paragraph{Acknowledgments}
The authors wish to acknowledge CSC – IT Center for Science, Finland, for computational resources.

\paragraph{Funding Statement}
This research was supported by the Research Council of Finland grant 341550.

\paragraph{Competing Interests}
None

\paragraph{Data Availability Statement}
Data used in this study is available at https://zenodo.org/records/12794391

\paragraph{Ethical Standards}
The research meets all ethical guidelines, including adherence to the legal requirements of the study country.

\paragraph{Author Contributions}
Conceptualization: A.P; I.M; A.N. Methodology: A.P; I.M.; Data visualisation: A.P. Writing original draft: A.P; I.M; A.N. All authors approved the final submitted draft.

\paragraph{Supplementary Material}
None

\bibliographystyle{unsrtnat}
\bibliography{references}

\end{document}